\documentclass[english]{article}

\usepackage{color}
\makeatletter
\usepackage[latin1]{inputenc}
\usepackage{graphicx}
\usepackage{geometry}
\makeatother

\usepackage{babel}

\begin{document}

\title{On the Nonrelativistic  Quantum-Mechanical Hamiltonian with $1/c^2$ terms.
Transverse  current-current interaction}

\author{Ladislaus Alexander B\'anyai \\
{\sl Institut f\"ur Theoretische Physik, Goethe-Universit\"at, Frankfurt am Main}}

\maketitle

\begin{abstract}
We extend the standard solid-state quantum mechanical Hamiltonian containing only Coulomb interactions between the charged particles by inclusion of $1/c^2$ terms representing (transverse) current-current interaction. For its derivation we use the classical formulation of Landau-Lifshitz, however consequently  in the Coulomb gauge retaining only the physical degrees of freedom. Our Hamiltonian does not coincide with the Darwin Hamiltonian. We emphasize the mathematical inconsistency in the derivation of this last Hamiltonian. We show, that the  quantized version of our Hamiltonian is equivalent to the  non-relativistic QED  considering only states without photons and retaining only terms of order $1/c^2$. The importance of this extended Hamiltonian  lies in the possibility  to distinguish external from internal magnetic fields. This aspect may be  relevant for theories of the Meissner effect.   
\end{abstract}

\maketitle
\section{Introduction}

The standard Hamiltonian of solid-state theory considers a system of electrons and ions interacting only through Coulomb forces. These are of zeroth order in  the $1/c$ expansion. Besides relativistic corrections, pure  electromagnetic interactions contribute already terms of order $1/c^2$  that might be important. The first classical Hamiltonian containing such terms was built up by Darwin \cite{Darwin} more than  hundred years ago. Later Landau-Lifshitz  \cite{Landau} re-derived the same result by the choice of a very special gauge. 

The interest for this approximation continues until today. We refer to earlier papers cited in the recent papers of Essen \cite{Essen1} \cite{Essen2}. This author also gave another derivation, instead of the $1/c$ expansion, just by neglecting the radiation effects in the Maxwell equations.  He underlined also the  importance of $1/c^2$  corrections in its quantum mechanical version for solid state  theory, particularly for the theory of superconductivity, as well as for the inclusion of ultra-relativistic effects. The application into plasma physics were discussed recently by  Krause et al.\cite{Krause}, aimed at the Vlasov equation. It is worth to mention also a paper by Bessonov \cite{Bessonov}, who found some strange solutions derived from the Darwin Lagrangian.    

The aim of this paper is to derive the correct extension of the solid-state Hamiltonian including terms of order $1/c^2$, since as we show, the Darwin Hamiltonian is misleading. Its derivation ignores some delicate, but essential aspects of the canonical formulation of the electromagnetic theory. We derive here the proper classical Hamiltonian  using  Landau-Lifshitz's \cite{Landau} way, but with the consequent use of the Coulomb gauge. The $1/c^2 $ terms we obtain describe a  (transverse) current-current interaction very similar to the density-density interaction of the Coulomb terms.  We prove here also that the quantized version of  our Hamiltonian coincides with the $1/c^2$ approximation of the non-relativistic Quantum Electrodynamics  (QED), restricted to the states without photons. This proves definitely the correctness of our construction.

An important feature of the extended Hamiltonian is the possibility to distinguish the internal and external magnetic fields. This might be important for a better understanding of superconductivity. Indeed, ideal diamagnetism (Meissner effect), consist in the compensation of the external magnetic field by the internal one. However, today's theories of superconductivity operate  within the frame of the standard Coulomb model of solid state and therefore here the magnetic field is just a self-consistent one.
 
\section{The classical $1/c^2$ Hamiltonian in the Coulomb gauge }

One can not formulate a Lagrangian theory of classical point-like charged particles interacting with the electromagnetic field due to the divergent self-interaction.( From the Lorentz force one has to omit the action of the field created by each charged particle on itself.) This impedes also the derivation of the appropriate Hamiltonian. Usually, one defines the classical or quantum mechanical theory of charged particles directly by a Hamiltonian including only Coulomb potentials without self-interacting terms.
Almost one hundred years ago Darwin \cite{Darwin} proposed a closed classical Lagrangian  for N point-like charges $e_i$ and mass $m_i$ ($i,j =1,\ldots N$) including terms up to order $1/c^2$ avoiding self.interaction and derived the corresponding Hamiltonian
\begin{eqnarray}
& &{\cal H}=\sum_{i}\frac{m_i}{2}\vec{p_{i}}^{2}+\sum_{i>j}\frac{e_i e_j}{|\vec{r_{i}}-\vec{r}_{j}|} \label{Darwin}
\\
& &-\sum_{i>j}\frac{e_i e_j}{2c^{2}m_i m_j|\vec{r}_{i}-\vec{r}_{j}|}\left[\vec{p}_{i}\cdot\vec{p}_{j}+(\vec{p}_{i}\cdot\vec{n}_{ij})(\vec{p}_{j}\cdot\vec{n}_{ij})\right] ,
\nonumber
\end{eqnarray}
where $\vec{n}_{ij}\equiv \frac{\vec{r}_i -\vec{r}_j}{|\vec{r}_i -\vec{r}_j| }$. 
His derivation is based on the expansion of the Li\'enard-Wiechert potentials to second order in $1/c$.  Jackson \cite{Jackson} in his derivation uses the Coulomb gauge, but is forced to make one more approximation to get the above result.
  
Landau-Lifshitz \cite{Landau} have shown that Eq. \ref{Darwin} actually implies a very unusual choice of gauge, not the usual Coulomb one. 
The choice of the gauge is however essential, since  the  physical magnetic field 
${\vec B}$, as well as the  photon in quantum electrodynamics (QED), have only two transverse degrees of freedom and only in the Coulomb  gauge (often called as the "physical" or "unitary" gauge) one is left just with these two degrees of freedom for the electromagnetic field. 
 The constraint on the vector potential (implied by Landau-Lifshitz's choice), after its elimination  through the velocities,  propagates on the velocities.
It is worth to recall here for example, that in the relativistic QED, in the Lorentz gauge, restrictions on the allowed physical states  have to be imposed to eliminate the longitudinal and temporal  photons! 
 
We follow here the way chosen by Landau-Lifshitz \cite{Landau} to construct a classical Hamiltonian up to terms of order $1/c^2$, however not in their choice of  gauge, but  in the Coulomb one. 
One  starts  with the Lagrangian of a single electron in an external field
(here in a non-relativistic approach!) produced by some external sources $\rho^{ext}$ and
$\vec{i}^{ext}$ 
\[
L(\vec{r,}\dot{\vec{r}})=\frac{m\dot{\vec{r}}^{2}}{2}-e\phi^{ext}(\vec{r},t)+
\frac{e}{c}{\cal\vec{A}}^{ext}(\vec{r},t)\dot{\vec{r}}\quad.
\]
In the Coulomb gauge
 \[\nabla {\cal\vec {A}}^{ext}(\vec{r},t)=0\]
the potentials are
\[
\phi(\vec{r},t)^{ext}=\int d\vec{x}\frac{\rho^{ext}(\vec{x},t)}{|\vec{r}-\vec{x}|};\quad\quad
{\cal \vec{A}}^{ext}(\vec{r},t)=\int d\vec{x}\frac{\vec{i}_{\bot}^{ext}(\vec{x},t-|\vec{r}-\vec{x}|/c)}{c|\vec{r}-\vec{x}|}\enspace ,
\]
where $\rho^{ex}({\vec x},t)$ is the external charge density, while   $\vec{i}_{\bot}^{ext}(\vec{x},t)$ is the external transverse ($\nabla\vec{i}_{\bot}^{ext}=0$)  current density

\[
\vec{i}_{\bot}^{ext}(\vec{x},t)\equiv\vec{i}^{ext}(\vec{x},t)+\frac{1}{4\pi}\nabla\int d\vec{x} '\frac{\nabla'\vec{i}^{ext}(\vec{x}',t)}{|\vec{x}-\vec{x}'|}\enspace .
\]
As Landau-Lifshitz do it, one has to expand the retarded current  density  in powers of $1/c$, however  here we need only the lowest approximation
due to the already existent  $1/c$  factor in the Lagrangian i.e.
\[
{\cal\vec{A}}^{ext}(\vec{r},t)\approx\int d\vec{x}\frac{\vec{i}_\bot^{ext}(\vec{x},t)}{c|\vec{r}-\vec{x}|}\enspace .
\]
If the source of the fields is a single point particle of charge $e'$  at $\vec{x}(t)$
having the velocity $\dot{\vec{x}}(t)$ then
 \[\rho^{ext}(\vec{x,}t)=e'\delta(\vec{x}-\vec{x}(t)),\qquad
\vec{i}^{ext}(\vec{x},t)=e'\dot{\vec{x}}(t)\delta(\vec{x}-\vec{x}(t))\enspace ,
\]
with
\[
\phi^{ext}(\vec{r},t)=\frac{e'}{|\vec{r}-\vec{x}(t)|} 
\]
and
\[{\cal\vec{A}}^{ext}(\vec{r},t)=
\frac{e' }{c}\left[\frac{\dot{\vec{x}}(t)}{|\vec{r}-\vec{x}(t)|}
-
\frac{1}{4\pi} \int d\vec{x}\frac{1}{|\vec{r}-\vec{x}|}\nabla \left(\dot{\vec{x}}(t)\nabla \frac{1}{|\vec{x}-\vec{x}(t)|}\right)\right]\enspace.
\]
Therefore the Lagrangian of the electron in the field of the another
electron, in this approximation, is

\begin{eqnarray*}
& &{\cal L}(\vec{r,}\dot{\vec{r;}}\vec{x,}\dot{\vec{x}})=\frac{m_i \dot{\vec{r}}^{2}}{2}-\frac{e e'}{|\vec{r}-\vec{x}(t)|}
\\
& &+\frac{e e'\dot{\vec{r}}}{c^2}\left[
\frac{\dot{\vec{x}}(t)}{|\vec{r}-\vec{x}(t)|}-\frac{1}{4\pi} \int d\vec{x}\frac{1}{|\vec{r}-\vec{x}|}\nabla \left(\dot{\vec{x}}(t)\nabla \frac{1}{|\vec{x}-\vec{x}(t)|}\right)\right]
 . 
\end{eqnarray*}

By generalization one obtains for a system of $N$ charged particles the total
Lagrange function
\begin{eqnarray*}
& &L\!=\!\sum_{i}\frac{m_i}{2}\vec{v_{i}}^{2}-\!\sum_{i>j}\frac{e_i e_j}{|\vec{r_{i}}-\vec{r}_{j}|}
\\
& &+\!\sum_{i>j}\frac{e_i e_j}{c^{2}}\vec{v}_i
\left[
\frac{\vec{v}_j }{|\vec{r}_i-\vec{r}_j|}-\frac{1}{4\pi} \int d\vec{x}\frac{1}{|\vec{r}_i-\vec{x}|}\nabla \left(\vec{v}_j \nabla \frac{1}{|\vec{x}-\vec{r}_j|}\right)\right] .
\end{eqnarray*}

Generally speaking, one has to use here Dirac's canonical  formalism \cite{Dirac1, Dirac2}, since due to the velocity dependent terms, there is a relationship between the canonical momenta, However, to lowest order in $1/c$ we have
\[
{\vec p}_i =\frac{\delta L}{\delta {\dot{\vec r}_i}}\approx m {\dot{\vec r}_i}
\]
and therefore (according to Landau-Lifshitz), we may still  remain in the frame of the standard canonical formalism. The resulting classical Hamiltonian is

\begin{eqnarray}
& & H =\sum_{i}\frac{\vec{p_{i}}^{2}}{2m_i}+\sum_{i>j}\frac{e_i e_j}{|\vec{r_{i}}-\vec{r}_{j}|}
\label{Hamclas}
\\
& &-\sum_{i>j}\frac{e_i e_j}{c^{2}m_i m_j}\vec{p}_i\left[
\frac{\vec{p}_j }{|\vec{r}_i-\vec{r}_j|}-\frac{1}{4\pi} \int d\vec{x}\frac{1}{|\vec{r}_i-\vec{x}|}\nabla \left(\vec{p}_j \nabla \frac{1}{|\vec{x}-\vec{r}_j|}\right)\right]\nonumber
\end{eqnarray}
including  $1/c^2$ terms. This Hamiltonian is not identical with the Darwin Hamiltonian
 Eq.\ref{Darwin}!

\section{Quantum-mechanical electron Hamiltonian with $1/c^2$ terms.\label{cur-cur}}
Now we may start to formulate directly a second quantized version of the theory  starting from the classical Hamiltonian of Eq.\ref{Hamclas}. For sake of simplicity we consider here a single sort of particles (electrons) of mass $m$ and charge $e$.

By introducing the charge and current densities:
\[
\rho(\vec{x})=\sum_i e\delta(\vec{x}-\vec{r}_i); \qquad 
\vec{i}(\vec{x})=\sum_i \frac{e}{m}\vec{p}_i\delta(\vec{x}-\vec{r}_i)
\]
one would be tempted to rewrite the classical Hamiltonia Eq.\ref{Hamclas}  as

\begin{equation}
\sum_{i}\frac{1}{2m}\vec{p_{i}}^{2}+\frac{1}{2} \int d\vec{x}\int d\vec{x}'
\frac{\rho(\vec{x})\rho(\vec{x}')}{|\vec{x}-\vec{x}')|}  -\frac{1}{2} \int d\vec{x}\int d\vec{x'}
\frac{\vec{i}_\bot(\vec{x})\vec{i}_\bot(\vec{x}')} 
{c^2|\vec{x}-\vec{x}'|}\enspace ,\label{Hsymb}
\end{equation}
where 
$\vec{i}_\bot (\vec{x})$ is the transverse part of the current density
\[
\vec{i}_{\bot}(\vec{r},t)\equiv\vec{i}(\vec{r},t)+\frac{1}{4\pi}\nabla\int d\vec{r}'\frac{\nabla'\vec{i}(\vec{r}',t)}{|\vec{r}-\vec{r}'|} \enspace .\label{c2-Hamil-cl}
\]
However, due to the divergent self-interaction of point-like classical particles this expression is not meaningful, even without the $1/c^2 $ terms.

The quantum-mechanical version of this Hamiltonian for a system of identical particles (here fermions) the problem is however milder. One cannot identify individual particles and therefore the self-interaction is at least not obvious and one may eliminate it partially in the second quantization formalism by considering a "normal ordering" of the operators in the Hamiltonian, as it was done also in the case of the Coulomb interaction. This ordering of the creation ans annihilation operators eliminates the interaction in states that contain less than two particles.
Therefore we  may proceed  with the second quantization formulation of the theory  directly  form the last symbolic expression Eq.\ref{Hsymb}. 

One has to introduce the second quantized  charge and the transverse part of the current density operators  expressed in terms of second quantized wave functions 
$\psi_{\sigma}({\vec x})$  for fermions with spin $1/2$    

 \[\rho(\vec{x})=e\sum_{\sigma=\pm1}\psi_{\sigma }^{+}(\vec{x})\psi_{\sigma}(\vec{x})\]
 
\[\vec{i}(\vec{x})=\frac{e}{2m}\sum_{\sigma=\pm1}\psi_{\sigma }^{+}(\vec{x})\frac{\hbar}{\imath}\nabla\psi_{\sigma}(\vec{x})+h.c. 
\] 

The resulting quantum mechanical Hamiltonian ${\bf H}$ in the second quantized formalism  looks then as
\begin{eqnarray}
{\bf H}=&-&\sum_{\sigma=\pm1}\int d{\vec x} \psi_{\sigma }^{+}(\vec{x})\frac{\hbar^2}{2m}\nabla ^2\psi_{\sigma}(\vec{x}) \label{Hamc2}
\\
&+&\frac{1}{2}\sum_{\sigma,\sigma'=\pm1}\int d{\vec x}\int d{\vec x}' \psi_{\sigma }^{+}(\vec{x})\psi_{\sigma' }^{+}(\vec{x}')
\frac{e^2}{|\vec{x}-\vec{x}')|}\psi_{\sigma'}(\vec{x}')\psi_{\sigma}(\vec{x}) \nonumber
\\
&-&\frac{1}{2} \int d\vec{x}\int d\vec{x'}
\frac{:\vec{i}_\bot(\vec{x})\vec{i}_\bot(\vec{x}'):} 
{c^2|\vec{x}-\vec{x}'|} \enspace . \nonumber 
\end{eqnarray}
The last term we did not write out explicitly, but just indicated the normal  ordering by the
 $:\ldots :$ symbols, since it is very lengthy and complicated in the coordinate space due to the additional integrals in the definition of the transverse part.  However, this term  has a simple expression in the discrete ${\vec k}$ - space basis (plane waves with periodical boundary conditions in a cube of volume $\Omega$). It looks explicitly as  
\begin{equation}
-\frac{e^{2}\hbar^{2}}{m^{2}c^2\Omega}\sum_{\sigma,\sigma'=\pm 1}\sum_{\vec{k},\vec{p},\vec{q}}\frac{2\pi}{q^{2}}\left(\vec{k}\vec{p}-\vec{q}\vec{k}\frac{1}{q^2}\vec{q}\vec{p}\right)a_{{\vec{k}},\sigma }^{+}a_{\vec{p},\sigma '}^{+}a_{\vec{p}+\vec{q},\sigma'} a_{\vec{k}-\vec{q},\sigma}
\end{equation}
This is the second quantized Hamiltonian of electromagnetic interacting electrons of order $1/c^2$. It includes a (transverse) current-current interaction. 

We still have to add the interaction with a classical external transverse vector potential $\vec {A}^{ext}(\vec {x},t)$.
According to the minimal rule, in the presence of $\vec{A}^{ext}$ one has to replace everywhere 
 $-\imath\hbar \nabla $ by 
$-\imath\hbar \nabla -\frac{e}{c}\vec{A}^{ext}$. This implies not only the modification of the kinetic energy term, but also the modification of the current density ${\vec i}_{\bot}(\vec{x})$   in the current-current term of the Hamiltonian. However, this will produce terms of order $1/ c^3 $ one may ignore.

One  has to  keep in mind nevertheless, that the current density  operator $\vec{j}(\vec x, t)$, whose average is of interest, must contain the diamagnetic term 
\[
\vec{j}(\vec{x},t)=\frac{e}{2m}\left(\psi^{+}(\vec{x})\left(\frac{\hbar}{\imath}\nabla-\frac{e}{c}\vec{A}^{ext}(\vec{x},t)\right)\psi(\vec{x},t)+h.c.\right)\enspace . \label{curdens}
\]

For sake of simplicity we discussed here only the electromagnetic  Hamiltonian of an electron system, assuming implicitly a uniform positive background. Actually it concerns also the basic non-relativistic Hamiltonian describing the electron-ion constituents of the solid state system and their e.m. interactions. It is known, however, that magnetic interactions of the spins, as well as the spin-orbit interaction,  pure relativistic QED contributions, are often relevant and have to  be included in solid-state theory. 
 
\section{Connection to the non-relativistic QED}

 One may look now at these results to see their meaning from the other side, namely form the point of view of the non-relativistic QED of electrons interacting with photons.  This
 Hamiltonian in the Coulomb gauge looks as 
\begin{eqnarray*}
 H^{QED} & = & \sum_{\vec{q},\lambda}\hbar\omega_{\vec{q}}b_{\vec{q},\lambda}^{+}b_{\vec{q},\lambda}
 \\
 &+&\sum_{\sigma=\pm1}\int d\vec{x}\psi_{\sigma}^{+}(\vec{x})\left[\frac{1}{2m}\left(\frac{\hbar}{\imath}\nabla-\frac{e}{c}\vec{A}(\vec{x})\right)^{2}\right]\psi_{\sigma}(\vec{x})\\
 & + & \sum_{\sigma,\sigma'=\pm1}\frac{1}{2}\int d\vec{x}\int 
 d{\vec x}'\psi_{\sigma}^{+}(\vec{x})\psi_{\sigma'}^{+}(\vec{x}')\frac{e^{2}}{|\vec{x}-\vec{x}'|}\psi_{\sigma'}(\vec{x}')\psi_{\sigma}(\vec{x})
\end{eqnarray*}
where the quantized transversal ($\nabla {\vec A}({\vec x})=0$ ) e.m. potential

\[
\vec{A}(\vec{x})=\sum_{\lambda=1,2}\sqrt{\frac{hc}{\Omega}}\sum_{\vec{q}}\frac{1}{\sqrt{|\vec{q}|}}\vec{e}_{\vec{q}}^{(\lambda)}
e^{\imath\vec{q}\vec{x}}\left(b_{\vec{q},\lambda}+b_{-\vec{q},\lambda}^{+}\right)
\]
was taken with periodical boundary conditions.

The photon frequency is $\omega_{\vec{q}}=c|\vec{q}|$, while the bosonic commutators are 
\[
\left[b_{\vec{q},\lambda},b_{\vec{q}',\lambda'}^{+}\right]=\delta_{\vec{q},\vec{q}}'\delta_{\lambda\lambda'}
\]
and the unit vectors $\vec{e}_{\vec{q}}^{(\lambda)}$ are orthogonal  to the wave vector  $\vec {q}$ and to each other  
\[
\vec{q}\vec{e}_{\vec{q}}^{(\lambda)}=0;\qquad \vec{e}_{\vec{q}}^{(\lambda)}\vec{e}_{\vec{q}}^{(\lambda')}=\delta_{\lambda\lambda'};
\qquad\vec{e}_{\vec{q}}^{(\lambda)}=\vec{e}_{-\vec{q}}^{(\lambda)};\qquad(\lambda,\lambda'=1,2) \enspace .
\]
Actually in the kinetic energy term of this Hamiltonian the product of the vector potentials ("seagull term") $\frac{e^2}{c^2}\int \psi^{+}\psi A A $ has to be normal ordered wit respect to the photon creation and annihilation operators!

As usual in many body theories of solid-state, here the non-relativistic QED, contrary to the fundamental relativistic QED, is understood as a cut-off theory, where the bare parameters coincide with the physical ones. 

Let us discuss this QED Hamiltonian on the subspace of unperturbed states containing only electrons and no photons (photon vacuum). Then, if we want to retain only contributions up to order $1/c^2$  we may omit from the beginning the (normal ordered) seagull" term $\frac{e^2}{c^2}\int \psi^{+}\psi A A $. Being itself of order $1/c^2$ it may have only even higher order non-vanishing matrix elements in this subspace.

\begin{figure}[h]
\begin{center}
\includegraphics[scale=0.08]{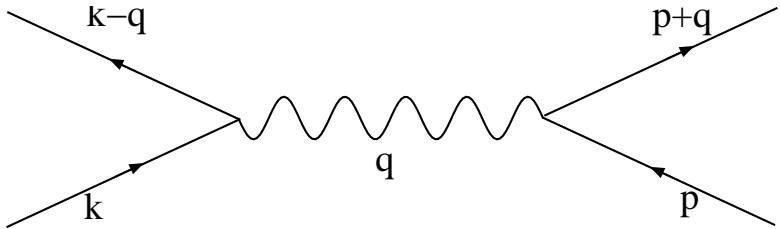}
\end{center}
\caption{The basic current-current graph in QED. }
\label{graph}
\end{figure}

Therefore, besides the standard Coulomb term one is left only with the photon-current interaction  $- \frac{1}{c}\int i_{\bot}A$. It appears in the S-matrix theory of adiabatic perturbations within  the considered subspace without photons only by Feynman diagrams constructed by the basic graph having four electron legs and two current-photon vertices connected by a photon propagator as shown in Fig.\ref{graph}.
After neglecting the term  $-\omega^2 /c^2 $  in the denominator of the photon  propagator in  Fourier space (i.e. eliminating corrections of higher order as $1/c^2$ already contained in the vertex parts) it looks as

\[\frac{1}{q^2}(\delta_{\mu,\nu}-\frac{q_\mu q_\nu }{q^2}); \qquad (\mu,\nu=1,2,3)\]
and one can convince oneself that this diagram of second order coincides with the first order diagram of the S matrix of the theory of the preceding Section. (The factor $-1/2$ in front of the current-current interaction term in  Eq.\ref{Hamc2} takes care of this last aspect!) 

Therefore, we may conclude, that the quantum mechanical $1/c^2$ electron Hamiltonian (more precisely the S-Matrix !) we obtained by the Landau construction in the Coulomb gauge, coincides indeed with the corresponding approximation in the non-relativistic QED.      

(The author  realized the above interpretation after consulting the first use of the term "current-current interaction"  by Holstein, Norton and Pincus \cite{Holstein} friendly suggested  by P. Kopietz.) 

\section{Relevance for the theory of superconductivity}
All theories of superconductivity are formulated in the frame of the lowest order in $1/c$ solid-state Hamiltonian describing Coulomb interacting electrons and ions. They are rather  successful in the phenomenological explanation of many experimental aspects. We have in mind mainly the Bardeen-Cooper-Schrieffer  \cite{BCS} (BCS) theory of the phase transition as being due to the anomalous correlation of electrons of opposite momenta and spins, caused by phonon exchange. The implementation of this idea is however performed within different further approximations. Among them the Bogolyubov-de Gennes equation \cite{Bogo-deGenn} is the most efficient, since by  considering a contact potential for the implementation of the BCS idea, this theory succeeds to formulate a description in the  real space and implicitly also at the boundaries. 
Another  succesful theory is due to Landau-Ginzburg \cite{Ginz-Land} based on original postulates without reference to a microscopical substrate.   

Despite all their successes neither of these theories is able to give a convincing explanation of the electromagnetic properties of superconductors. Here we do not think  about the hopeless proof of current flow without dissipation. This would be a task for the theory of open systems. Unfortunately, this theory, in spite of the big progresses of the last 60 years, is not yet able to answer such questions. (See a good review of the state of art by Spohn \cite{Spohn}.) 

We mean here just the understanding of already the simplest phenomenon, the Meissner effect. Obviously, the expulsion of the magnetic field from a superconductor in equilibrium is just the ideal diamagnetism, the internal magnetic field compensating the external one. Although, seemingly one succeeds with today's theories to show the failure of the penetration of the magnetic field, one remains confused about the underlying physical mechanism, since within the zeroth order in $1/c$ description one can make no distinction between the external and internal magnetic fields. All today's theories have to resort to a self-consistent treatment of the magnetic field. A distinction between the external and internal magnetic fields is possible either within the QED or its $1/c^2$ approximation described in this paper. 
To formulate a theory of superconductivity within this new frame is not at all simple, but desirable.

\section{Conclusions}
We have built up the correct extension of the basic Hamiltonian of classical interacting point-like charged particles  in order to include electromagnetic effects of order $1/c^2$ using  Landau-Lifshitz's \cite{Landau} construction, however in the adequate Coulomb gauge. We have shown also, that its quantized  version is equivalent to ignoring higher than $1/c^2$ corrections in the QED Hamiltonian and   restricting the theory to the subspace of free electron-ion  states without photons. This implies omitting the "sea-gull" vertex and the retardation in the photon propagator. We stressed also the importance of the inclusion of current-current interaction for a convincing theory of the Meissner effect. We hope, that this extended $1/c^2$ theory may be relevant also in  other fields of solid-state theory, as well as, in the classical version, for the plasma theory.

\vspace{1cm}

{\bf \large{Acknowledgments}}
\vspace{0.5cm}

The author is  deeply indebted to P. Gartner, M. Bundaru, V. B\^arsan, P. Kopietz and H. Haug for useful discussions on this topic.

\end{document}